\documentclass[aps,prl,reprint,superscriptaddress,nofootinbib]{revtex4-1}
\usepackage{hyperref} 

\usepackage[centertags,sumlimits,intlimits,namelimits,reqno]{amsmath}
\usepackage{graphicx}

\newcommand {\be}{\begin{eqnarray}}
\newcommand {\ee}{\end{eqnarray}}
\newcommand {\beq}{\begin{equation}}
\newcommand {\eeq}{\end{equation}}
\newcommand {\bp} {\begin{pmatrix}}
\newcommand {\ep} {\end{pmatrix}}

\newcommand{\E}{\mathcal{E}}

\newcommand{\bra}[1]{\langle {#1}|}
\newcommand{\ket}[1]{|{#1}\rangle}

\newcommand{\ketbra}[2]{\ket{#1}\negmedspace\bra{#2}}

\def\E{{\mathcal E}}
\def\Z{{Z}}
\def\X{{X}}
\def\Y{{Y}}

\begin{document}

\title{
Demonstration of sufficient control for two rounds of quantum error correction in a solid state ensemble quantum information processor
}

\author{
Osama Moussa
}
\email{omoussa@iqc.ca}
\affiliation{
Institute for Quantum Computing, 
University of Waterloo, Waterloo, ON, N2L 3G1, Canada.
}
\affiliation{
Department of Physics and Astronomy, 
University of Waterloo, Waterloo, ON, N2L 3G1, Canada.
}

\author{
Jonathan Baugh
}
\affiliation{
Institute for Quantum Computing, 
University of Waterloo, Waterloo, ON, N2L 3G1, Canada.
}
\affiliation{
Department of Chemistry, 
University of Waterloo, Waterloo, ON, N2L 3G1, Canada.
}
\author{
  Colm A. Ryan
}
\affiliation{
Institute for Quantum Computing, 
University of Waterloo, Waterloo, ON, N2L 3G1, Canada.
}
\affiliation{
Department of Physics and Astronomy, 
University of Waterloo, Waterloo, ON, N2L 3G1, Canada.
}
\author{
Raymond Laflamme
}
\affiliation{
Institute for Quantum Computing, 
University of Waterloo, Waterloo, ON, N2L 3G1, Canada.
}
\affiliation{
Department of Physics and Astronomy, 
University of Waterloo, Waterloo, ON, N2L 3G1, Canada.
}
\affiliation{
Perimeter Institute for Theoretical Physics, Waterloo, ON, N2J 2W9, Canada.
}

\begin{abstract}
We report the implementation of a 3-qubit quantum error correction code (QECC) on a quantum information processor realized by the magnetic resonance of Carbon nuclei in a single crystal of Malonic Acid. The code corrects for phase errors induced on the qubits due to imperfect decoupling of the magnetic environment represented by nearby spins, as well as unwanted evolution under the internal Hamiltonian. We also experimentally demonstrate sufficiently high fidelity control to implement two rounds of quantum error correction. This is a demonstration of state-of-the-art control in solid state nuclear magnetic resonance, a leading test-bed for the implementation of quantum algorithms.  
\end{abstract}


\maketitle

{\bf Introduction --}
One of the crucial requirements~\cite{Div00a} for universal quantum information processing (QIP) is the ability to protect the fragile quantum information during computation -- either by encoding the information in subspaces of the system's Hilbert space where it is protected from degradation by noise, or by an active scheme that detects and rectifies errors continuously or periodically. 
This latter, active, technique has been experimentally realized in liquid-state-NMR~\cite{CPM+98,LVZ+99,KLMN01}  and ion-trap~\cite{CLS+04} implementations of a quantum information processor. In each of these cases, a quantum error-correction code (QECC) was used to protect against the particular errors present in the respective systems; and it was shown that even with imperfect encoding and recovery operations, employing quantum error correction is advantageous. Obviously, there is a limit to how many control errors can be tolerated before they overwhelm the error correction protocol. Thus, the ability to demonstrate error correction is a highly relevant benchmark of high-fidelity coherent control. 

A natural question to ask is, whether one has high-enough-fidelity control to perform multiple rounds of error correction, as would be required in a realistic computation. Of course, to usefully perform multiple rounds, one needs a fresh supply of sufficiently pure {ancill\ae} to ensure that entropy flows in the proper direction. However, \emph{assuming} we have a fresh supply of {ancill\ae}, is it possible, with the current level of control, to perform meaningful multiple rounds of quantum error correction? 

In this manuscript, we report on the implementation of a three-qubit QECC that corrects phase errors induced by the environment, in a single-crystal solid-state NMR system. We also devise a way to experimentally determine the entanglement fidelity of multiple back-to-back rounds of error correction, and use it to determine the entanglement fidelity of two rounds of the 3-bit phase code. In light of recent work on the characterization and control of such systems~\cite{LGO+03,BMR+06,MS06}, as well as state initialization~\cite{BMR+05,RMB+08}, this current work signifies an advancement of one of the leading test-beds for QIP ideas. 
The rest of this manuscript is organized as follows. First, we describe the solid state NMR system, and the sources of noise affecting the qubits. We then describe the QECC implemented, and show the results from one and two rounds.

{\bf System and error models --} 
Building on the success of liquid-state NMR as a test bed of QIP ideas, 
Solid-state NMR
systems offer~\cite{CLK+00,BMR+06}
intrinsically larger couplings, longer coherence times, the ability to pump
entropy out of the system of interest into a spin bath
and the potential for much higher initial polarizations. This comes at the cost of a more complicated Internal Hamiltonian, which makes the system harder to control in practice.

The computational register under investigation is an ensemble of molecular nuclear spins in
 a macroscopic single crystal of Malonic Acid (C$_3$H$_4$O$_4$). A small fraction ($\sim 3\%$) of the molecules are triply labeled with (spin-$\tfrac{1}{2}$) $^{13}$C to form an ensemble of processor molecules, spatially buffered from one another by molecules of the same compound but with  natural abundance ($\sim1\%$) Carbon nuclei. During computation, the processors are decoupled from the 100\% abundant spin-$\tfrac{1}{2}$ protons in the crystal by applying a decoupling pulse sequence~\cite{FKE00} to the protons.  
For this 3-qubit register, (dephasing) noise comes in the following forms:
\begin{itemize}
\item {\bf Coherent phase errors} due to pulse implementation errors, phase transients, or unwanted evolution under the natural Hamiltonian (e.g. under the Zeeman term.) These are unitary errors (that cause no loss of coherence) and can therefore be inverted if tracked properly, but in case tracking that evolution is not possible, quantum error correction becomes a valuable tool.
 
\item {\bf Incoherent phase errors} due to Zeeman-shift dispersion or other inhomogeneities. The loss of coherence is over the ensemble; each member of the ensemble sees a different value for some coupled classical degree of freedom. Errors of this nature have been dealt with using refocussing techniques (e.g. spin echo), or by carefully designing the control fields to generate the same evolution over the ensemble distribution of the inhomogeneous parameter. Quantum error correction can be used with, or in lieu of, these other techniques to improve robustness to ensemble errors.
 
\item {\bf Decoherent phase errors} due to a coupling between the system of interest and environment  --an uncontrollable quantum degree of freedom-- and loss of coherence occurs when this environment is traced over after the interaction. 

\end{itemize}

\textbf{3-bit code -- }
The 3-bit repetition code was introduced by Shor~\cite{Shor95} as part of a 9-qubit code that is able to protect against an arbitrary single qubit error. The phase variant~\cite{Bra+96} of the 3-bit repetition code encodes a single qubit in three qubits as follows:
\beq
\ket{0} \to \ket{\bar{0}}=\ket{{+++}}\,,
\ket{1} \to \ket{\bar{1}}=\ket{{---}}\,;
\eeq
where $\ket{\pm}=\ket{0}\pm\ket{1}$, and the logical basis are denoted by $\{\ket{\bar{0}},\ket{\bar{1}}\}$. In the stabilizer formalism~\cite{Got97a,Pou05}, the stabilizer group generators for this code are $\{ XXI , IXX\}$. This code can be employed to correct for various sets of errors by choosing different decoding circuits -- for this work, we design the decoding to correct for errors generated by the set $\E=\{ZII, IZI, IIZ, III\}$.
That is to say, with the same decoding circuit the code corrects
a coherent selective phase rotation on one of the qubits,
\begin{equation*}
\Z_1^\theta
:=
e^{-i\ \theta/2\ \Z II}
=
\cos (\theta/2) \ III - i \sin (\theta/2)\ \Z II\,,
\end{equation*}
and/or the dephasing map on one of the qubits, due to incoherent/decoherent errors:
\begin{equation*}
 \Lambda_\theta (\rho) = \cos^2 (\theta) I \rho I + \sin^2 (\theta) \Z \rho \Z\,.
 \end{equation*}

\begin{figure}[h]
\includegraphics[scale=.7]{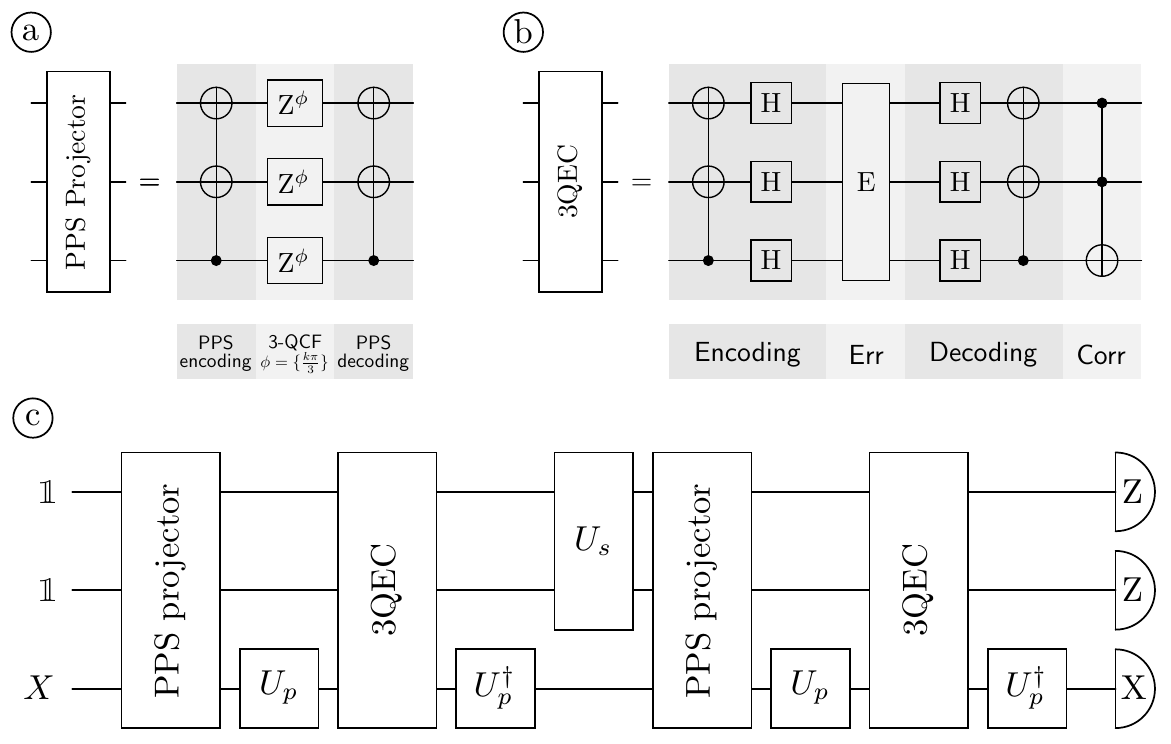}
\caption{\label{figqeccircuit}  Shown are the implemented quantum circuits for (a)  Labeled-pseudopure-state (PPS) preparation procedure: a triple-quantum-coherence-filte (3QCF) is conjugated by a unitary operation that encodes (and decodes) the labeled pseudopure state $\ketbra{00}{00}X$ in the triple quantum coherence $\ketbra{000}{111}+\ketbra{111}{000}$; (b) The implemented quantum circuit of a three-qubit QECC, showing the encoding,
decoding, and error-correction steps. The top two qubits are initialized to the $\ket{00}$ state, and the bottom qubit carries the information to be encoded. After the decoding and correction operations, the bottom qubit is restored to its initial state, while the top two qubits carry information about which error had occured; and (c) The procedure for two rounds:  $U_p$ prepares X,Y, or Z inputs, and $U_s=\{ II, XI, IX, XX\}$ toggles between the different syndrome subspaces. i.e. The experiment is repeated four times, cycling through the different $U_s$, and then the results are added, similar to a standard phase cycling procedure. }
\end{figure}

A quantum circuit that accomplishes~\cite{Bra+96} the encoding,
decoding, and error-correction steps is shown in Figure~\ref{figqeccircuit}-b. 
The encoding process takes a qubit in the state $\alpha \ket{0} + \beta \ket{1}$, as well as two ancillary qubits prepared in the $\ket{00}$ state, and outputs the 3-qubit encoded state $\alpha \ket{{+++}} + \beta \ket{{---}}$. After the error channel, the recovery process decodes the state on the information-carrying-qubit, and the other qubits carry syndrome information about the errors that have occurred. The nondegeneracy of the code implies that each of the error basis, in $\E$, will leave a particular signature. It is straightforward to show that syndromes 00, 10, 01, and 11 correspond to the occurrence of errors $III$, $\Z II$, $I\Z I$, and $II\Z$, respectively.

The figure of merit used herein to judge the performance of the code is the \emph{entanglement fidelity}~\cite{Sch96a}. In particular, we use the expression for the single-qubit average entanglement fidelity, which is experimentally accessible by measuring the fraction of surviving signal given input states $\X$,   $\Y$, and $\Z$~\cite{FVH+02}.

\begin{figure}
\hspace{-.4in}
\begin{minipage}{0.4\linewidth}
\includegraphics[scale=.18]{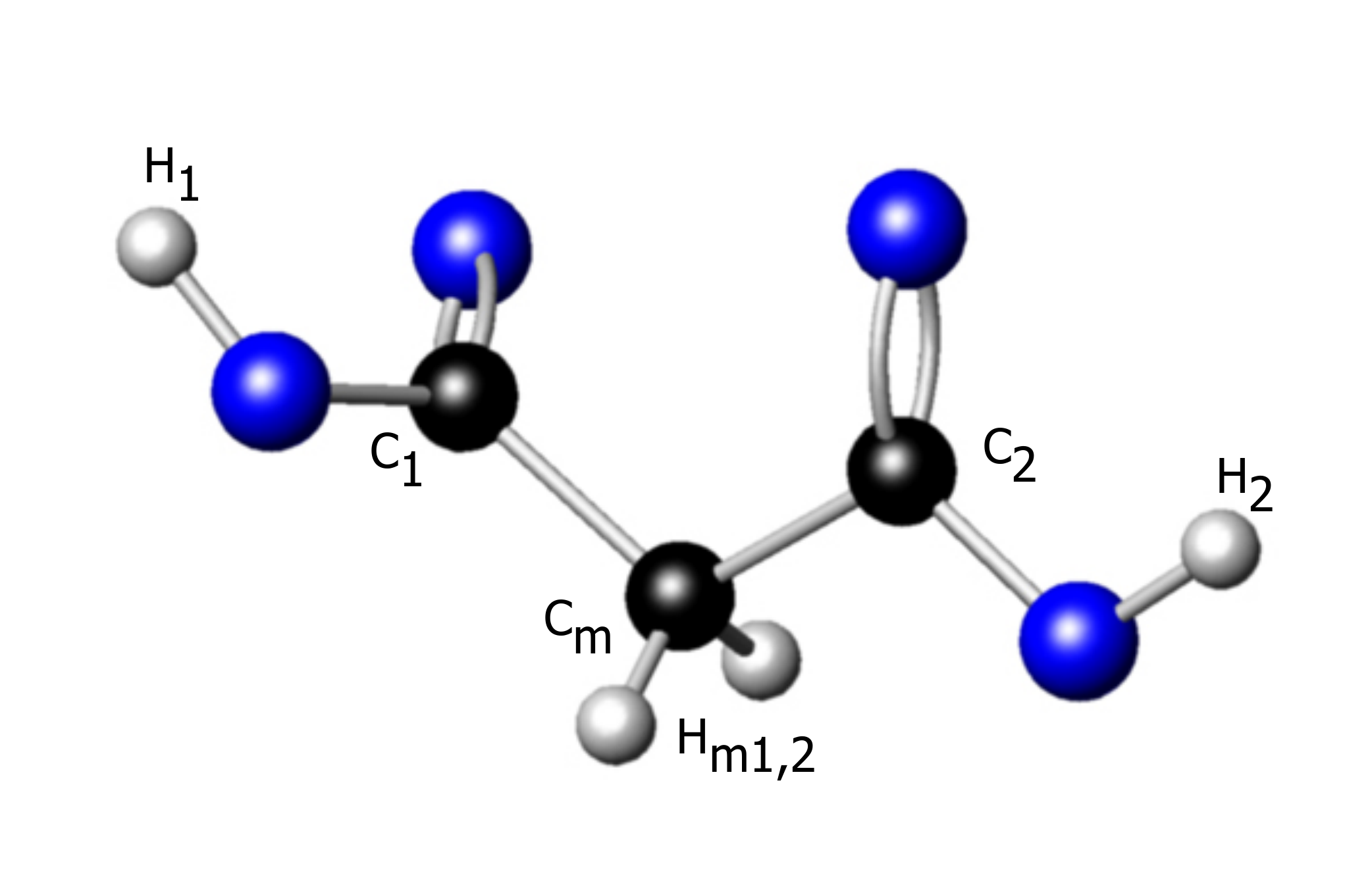}
\end{minipage}
\hspace{0.5cm}
\begin{minipage}{0.35\linewidth}
 \begin{small}
\begin{tabular}{|c|c|c|c|}
\hline
kHz&C$_1$&C$_2$&C$_m$\\
\hline
C$_1$&6.380&0.297&0.780\\
\hline
C$_2$&-0.025&-1.533&1.050\\
\hline
C$_m$&0.071&0.042&-5.650\\
\hline
\end{tabular}
\end{small}
\end{minipage}\\
\includegraphics[width=2.8in]{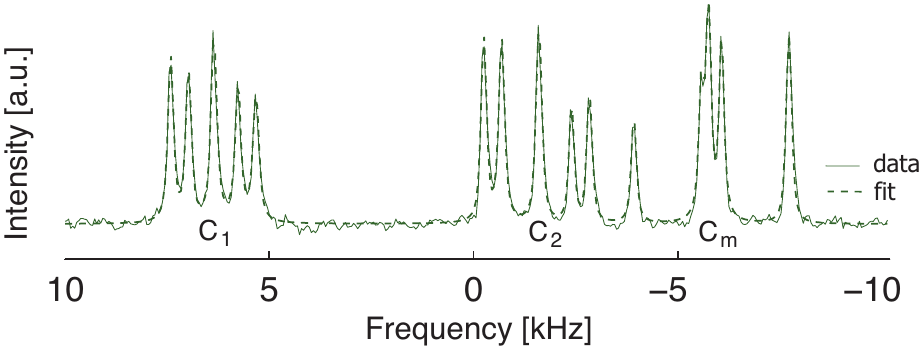}
\caption{\label{figmalonic} Malonic acid (C$_3$H$_4$O$_4$) molecule and Hamiltonian parameters (all values in kHz).  Elements along the diagonal represent chemical shifts, $\omega_i$, with respect to the transmitter frequency (with the Hamiltonian $\sum_i \pi \omega_i \Z_i$). Above the diagonal are dipolar coupling constants ($\sum_{i<j} \pi D_{i,j} (2\ \Z_i\Z_j - \X_i\X_j - \Y_i\Y_j$), and below the diagonal are J coupling constants, ($\sum_{i<j} \tfrac{\pi}{2} J_{i,j} (\Z_i\Z_j + \X_i\X_j + \Y_i\Y_j$).  An accurate natural Hamiltonian is necessary for high fidelity control and is obtained from precise spectral fitting of (also shown) a proton-decoupled $^{13}$C spectrum following polarization-transfer from the abundant protons. The central peak in each quintuplet is due to natural abundance $^{13}$C nuclei present in the crystal at $\sim 1\%$. (for more details see~\cite{BMR+06,RMB+08} and references therein.) }
\end{figure}

\textbf{Experiment -- } The experiments were performed in a static field of $7.1$T using a purpose-built NMR probe. Shown in Figure~\ref{figmalonic} is a proton-decoupled $^{13}$C spectrum, following polarization-transfer from the abundant protons, for the particular orientation of the crystal used in this experiment. A precise spectral fit gives the Hamiltonian parameters (listed in the inset table in Figure~\ref{figmalonic}), as well as the free-induction dephasing times, $T_2^*$, for the various transitions; these average at $\sim2$ms. The dominant contribution~\cite{BMR+06} to $T_2^*$ is Zeeman-shift dispersion, which is largely refocused by the control pulses. Other contributions are from intermolecular $^{13}$C-$^{13}$C dipolar coupling and, particularly for C$_m$, residual interaction with neighboring protons due to imperfect decoupling.
The carbon control pulses are numerically optimized to implement the required unitary gates using the GRAPE~\cite{KRK+05} algorithm. A typical pulse is $1$ms long, and is designed~\cite{RNL+08} to have an average Hilbert-Schmidt fidelity of $99.8\%$ over appropriate distributions of Zeeman-shift dispersion and control-fields inhomogeneity.

The system is initially prepared in the labelled pseudopure state (PPS)~\cite{CPH98a,KCL98,KLMT00} --expressed in the product operator formalism-- $\rho_i=\tfrac{1}{8}[III+\epsilon (I+\Z)(I+\Z)\X]$, where $\epsilon \sim 10^{-5}$. The completely mixed component is ignored for the rest of the discussion, as it does not participate in the unital dynamics. 
This preparation is achieved by control pulses and phase cycling (temporal averaging), and can be thought of as a projection along $(I+\Z)(I+\Z)\X$. 

As shown in Figure~\ref{figqeccircuit}-a, the phase cycling actuates a \emph{triple-quantum-coherence filter} (3QCF)~\cite{PSE82} by exploiting the $n$-proportional phase acquisition of $n$-coherence quantum states under $\Z$-rotation. And conjugating the 3QCF with transformations that encode (and later decode) the $(I+\Z)(I+\Z)\X$ coherence in the triple-quantum-coherence, $\ketbra{000}{111}+\ketbra{111}{000}$, realizes an effective projector on the labelled-PPS.

\begin{figure}
\begin{center}
\includegraphics[scale=.65, trim=5 0 0 0, clip]{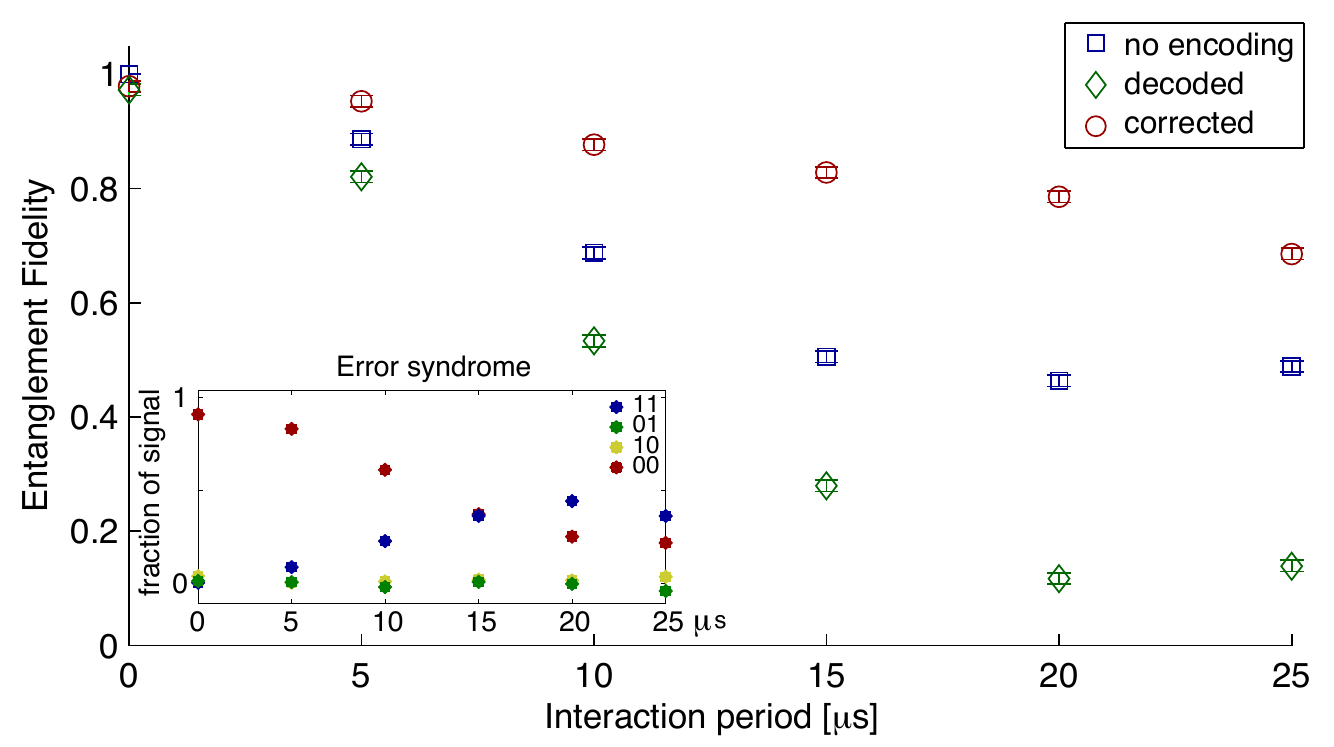}
\caption{\label{figfullHC}Experimentally determined entanglement fidelities for  
unencoded (blue squares), before (green diamonds), and after (red circles) the correction step of one round of QEC. After encoding, a variable delay is implemented before the recovery process. During the delay, the system evolves under the heteronuclear and homonuclear terms in the natural Hamiltonian. The non-montonicity in the unencoded and decoded data are 
indicative of the presence of a coherent error. The inset shows the intensities measured for the different syndromes; the dominant error is a phase rotation on the 
bottom qubit (C$_m$). }
\end{center}
\end{figure}

We first examine the performance of the 3-bit phase code under the natural evolution of the system: between the encoding and recovery operations, the system is 
left to evolve unobstructed under the full natural Hamiltonian, both homonuclear 
and heteronuclear parts. Exaggerated as it is, this is a useful test of the code's ability to correct for coherent errors from uncertainties in the natural Hamiltonian, or imperfect decoupling of the magnetic environment. The experimentally determined entanglement fidelities 
are shown in Figure~\ref{figfullHC}, demonstrating the advantage of quantum error correction. 
The syndrome information (inset in Figure~\ref{figfullHC}) indicate that the dominant 
phase error is on the methylene carbon, C$_m$. The non-montonicity of the unencoded and 
decoded data indicate that the error is, at least partially, coherent. However, full simulation of the dynamics of the carbon subsystem suggest a longer timescale for the homonuclear coherent effects. Moreover, the timescale of the revival of the signal is consistent with the coupling strength between C$_m$ and the methylene protons, which leads us to conclude that this coupling is responsible for the non-motonicity in the entanglement-fidelity decay. This conclusion is further supported by the following results, where this coupling is partially averaged using a heteronuclear-dipolar decoupling pulse sequence.

\begin{figure}
\includegraphics[scale=.5, trim=2 0 0 0, clip]{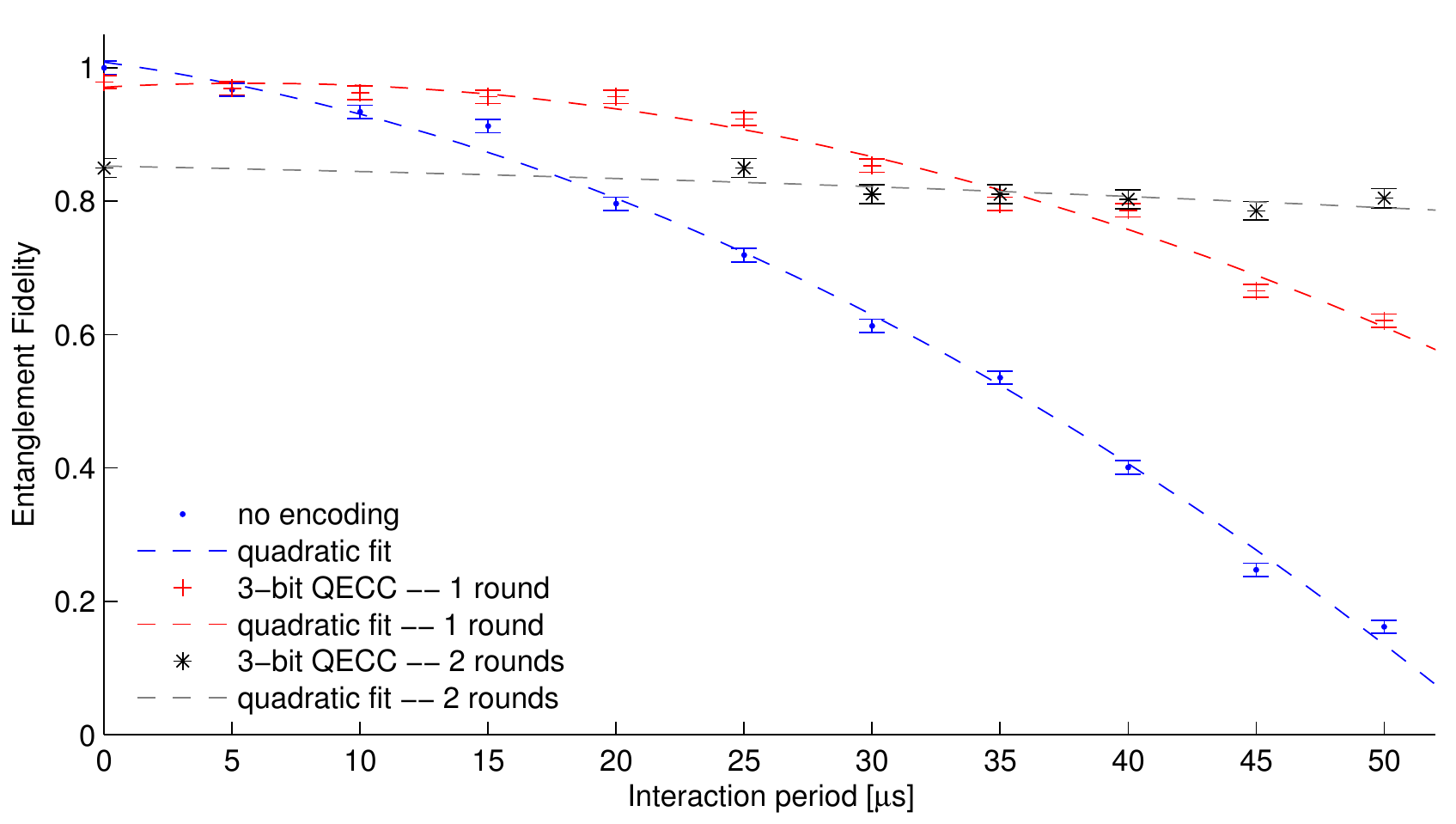}
\includegraphics[scale=.36, trim=-14 0 0 0, clip]{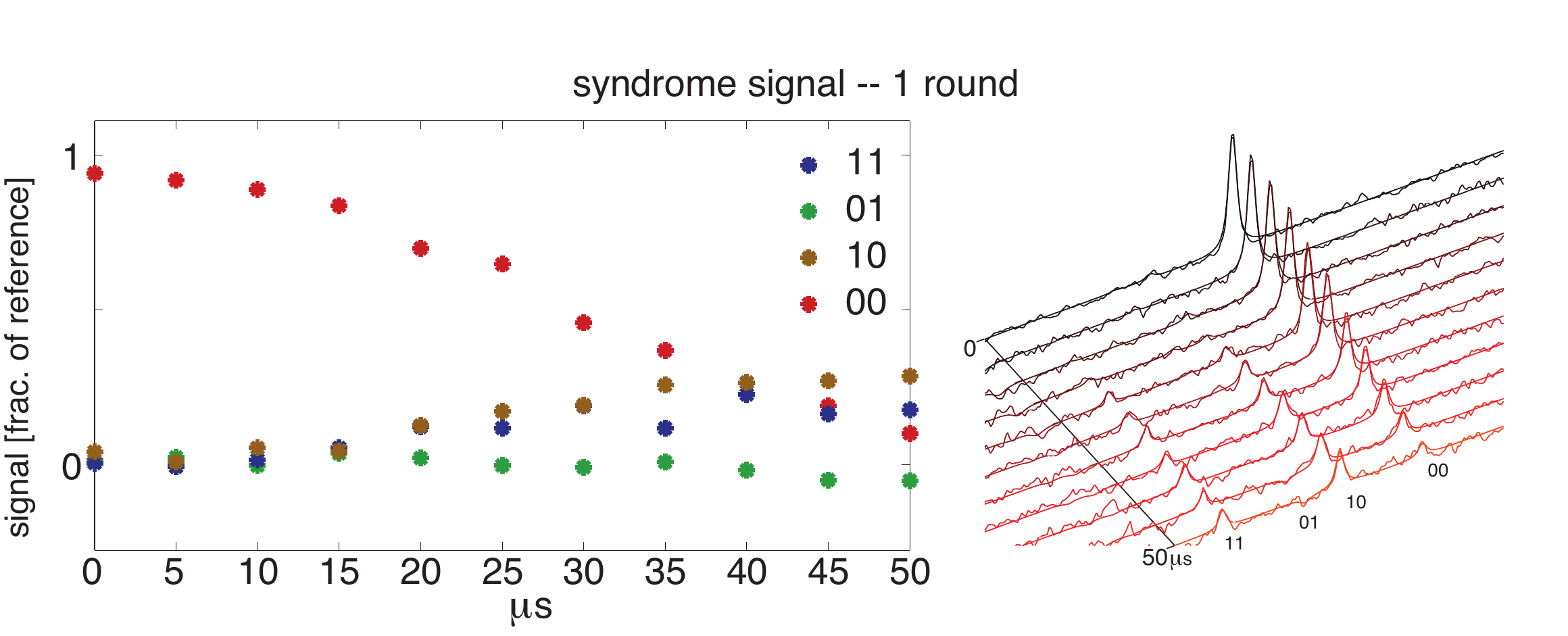}
\caption{\label{figtworounds}Summary of experimental results for the \emph{partial decoupling} map: the system evolves under the natural Hamiltonian as well as 70kHz decoupling fields that partially modulate the heteronuclear interactions (between the carbons and protons.) Shown (in top figure) are the single-qubit entanglement fidelities in the cases where no encoding is employed (blue dots); or one round of the 3-bit code (red crosses); or two rounds of the 3-bit code (black asterisks), where the interaction interval is split to two equal intervals. The dashed lines are quadratic fits to the data, and are included to guide the eye. Also shown (bottom) is the signal after one round of error correction as distributed over the various error-syndrome subspaces. In this case, the dominant errors are phase flips on the top and bottom qubits, which are encoded on C$_1$ and C$_m$, respectively.}
\end{figure}

Next, the 3-bit phase code is employed to protect a single qubit against errors from evolution under the natural Hamiltonian of the carbon subsystem as well as residual heteronuclear couplings between the carbons and protons due to partial decoupling of the protons using the SPINAL64 sequence~\cite{FKE00}  at amplitude of 70kHz. 
From the syndrome information shown in Figure~\ref{figtworounds},  the major contributions are from phase rotations on C$_1$ and C$_m$. This is to be expected, since, in this orientation, and in this rotating frame, the Zeeman shifts of these two spins are the dominant terms in the internal  Hamiltonian. 

\textbf{Two rounds -- } We devise a way to experimentally determine the entanglement fidelity of multiple rounds of error correction and use it to experimentally determine the entanglement fidelity of two rounds of the 3-bit phase code. After the first round of error correction, the surviving polarization from the various input states is distributed over the various subspaces of the Hilbert space corresponding to the various syndromes. For the second round, for each syndrome, we project into the subspace of the syndrome and perform error correction in that subspace, and then sum over all possible syndromes. 
For each syndrome, this projection is implemented as a transformation (denoted by $U_s$ in Figure ~\ref{figqeccircuit}-c) that swaps the information in that subspace with the subspace where the ancillas are in (I+\Z)(I+\Z) --or $\ketbra{00}{00}$-- and then projecting unto the latter subspace using the same protocol for initial state preparation.
The quantum circuit describing the protocol is shown in Figure~\ref{figqeccircuit}-c, and the experimental results are shown in Figure~\ref{figtworounds}. 

The results show that, for long interaction intervals, there is an advantage to performing two rounds of error correction with our current level of control. The initial drop (at zero interaction interval) in the experimentally determined two-round entanglement fidelity is mainly due to the projection operation, which is not needed if pure ancill\ae are available. 

The scheme requires a number of experiments that grows as $s^{m-1}$, where $s$ is the number of possible nondegenerate syndromes of the code, and $m$ is the number of rounds of correction performed. In this sense, the applicability of the scheme is very limited, but it is sufficient for our purposes.

\textbf{Conclusion -- }
We were able to demonstrate the advantage of performing quantum error correction to protect against relevant, naturally occurring phase errors --coherent, incoherent and decoherent-- that arise in a solid-state system. We have shown that this is possible by achieving state-of-the-art control on a 3-qubit system. Moreover, we have shown that with these control fidelities, multiple rounds of QEC are possible.  This is particularly significant in a system where it has been previously shown that entropy can be efficiently removed from the system of interest to the environment~\cite{BMR+05,RMB+08}. 

 \textbf{Acknowledgements --} 
O.M. would like to acknowledge helpful discussions with D. G. Cory, and technical support from M. Ditty. This research was supported by NSERC, CIFAR, and the Premier Discovery Awards.

\end{document}